\documentclass[]{aastex631}

\def\arcsec{{\rm\thinspace arcsec}}
\def\cm{{\rm\thinspace cm}}

\def\g{{\rm\thinspace g}}

\def\G{{\rm\thinspace G}}

\def\K{{\rm\thinspace K}}
\def\keV{{\rm\thinspace keV}}

\def\m{{\rm\thinspace m}}

\def\Mpc{{\rm\thinspace Mpc}}

\def\pc{{\rm\thinspace pc}}

\def\s{{\rm\thinspace s}}

\def\ks{{\rm \thinspace ks}}
\def\arcsec{\hbox{$^{\prime\prime}$}}

\shorttitle{X-ray Polarization Limits for Cen A}
\shortauthors{Ehlert et al.}
\graphicspath{{./}{figures/}}
\begin{document}

\title{Limits on X-ray Polarization at the Core of Centaurus A as Observed with the Imaging X-ray Polarimetry Explorer}

\correspondingauthor{Steven R. Ehlert}
\email{steven.r.ehlert@nasa.gov}
\author[0000-0003-4420-2838]{Steven R. Ehlert}
\affiliation{NASA Marshall Space Flight Center, Huntsville, AL 35812, USA}  
\author[0000-0003-1074-8605]{Riccardo Ferrazzoli}
\affiliation{INAF Istituto di Astrofisica e Planetologia Spaziali, Via del Fosso del Cavaliere 100, 00133 Roma, Italy}
\author[0000-0002-2055-4946]{Andrea Marinucci}
\affiliation{Agenzia Spaziale Italiana, Via del Politecnico snc, 00133 Roma, Italy}
\author[0000-0002-6492-1293]{Herman L. Marshall}
\affiliation{MIT Kavli Institute for Astrophysics and Space Research, Massachusetts Institute of Technology, 77 Massachusetts Avenue, Cambridge, MA 02139, USA}
\author[0000-0001-9815-9092]{Riccardo Middei}
\affiliation{Space Science Data Center, Agenzia Spaziale Italiana, Via del Politecnico snc, 00133 Roma, Italy}
\affiliation{INAF Osservatorio Astronomico di Roma, Via Frascati 33, 00040 Monte Porzio Catone (RM), Italy}
\author[0000-0001-6897-5996]{Luigi Pacciani}
\affil{Istituto di Astrofisica e Planetologia Spaziali - Istituto Nazionale di Astrofisica (IAPS-INAF), Via Fosso del Cavaliere, 100 - I-00133, Rome, Italy}
\author[0000-0003-3613-4409]{Matteo Perri}
\affiliation{Space Science Data Center, Agenzia Spaziale Italiana, Via del Politecnico snc, 00133 Roma, Italy}
\affiliation{INAF Osservatorio Astronomico di Roma, Via Frascati 33, 00040 Monte Porzio Catone (RM), Italy}
\author[0000-0001-6061-3480]{Pierre-Olivier Petrucci}
\affiliation{Universit\'{e} Grenoble Alpes, CNRS, IPAG, 38000 Grenoble, France}
\author[0000-0002-2734-7835]{Simonetta Puccetti}
\affiliation{Agenzia Spaziale Italiana, Via del Politecnico snc, 00133 Roma, Italy}
\author[0000-0003-1340-5675]{Thibault Barnouin}
\affiliation{Universit\'{e} de Strasbourg, CNRS, Observatoire Astronomique de Strasbourg, UMR 7550, 67000 Strasbourg, France}
\author[0000-0002-4622-4240]{Stefano Bianchi}
\affiliation{Dipartimento di Matematica e Fisica, Universit\`{a} degli Studi Roma Tre, Via della Vasca Navale 84, 00146 Roma, Italy}
\author[0000-0001-9200-4006]{Ioannis Liodakis}
\affiliation{Finnish Centre for Astronomy with ESO, University of Turku, Vesilinnantie 5, 20014 Turku, Finland}
\author{Grzegorz Madejski}
\affiliation{SLAC National Accelerator Laboratory 2575 Sand Hill Road Menlo Park, CA 94025, USA}
\author[0000-0003-4952-0835]{Fr\'{e}d\'{e}ric Marin}
\affiliation{Universit\'{e} de Strasbourg, CNRS, Observatoire Astronomique de Strasbourg, UMR 7550, 67000 Strasbourg, France}
\author[0000-0001-7396-3332]{Alan P. Marscher}
\affiliation{Institute for Astrophysical Research, Boston University, 725 Commonwealth Avenue, Boston, MA 02215, USA}
\author[0000-0002-2152-0916]{Giorgio Matt}
\affiliation{Dipartimento di Matematica e Fisica, Universit\`{a} degli Studi Roma Tre, Via della Vasca Navale 84, 00146 Roma, Italy}
\author[0000-0002-0983-0049]{Juri Poutanen}
\affiliation{Department of Physics and Astronomy, 20014 University of Turku, Finland}
\affiliation{Space Research Institute of the Russian Academy of Sciences, Profsoyuznaya Str. 84/32, Moscow 117997, Russia}
\author[0000-0002-7568-8765]{Kinwah Wu}
\affiliation{Mullard Space Science Laboratory, University College London, Holmbury St Mary, Dorking, Surrey RH5 6NT, UK}
\author{Iv\'{a}n Agudo}
\affiliation{Instituto de Astrof\'isica de Andaluc\'ia—CSIC, Glorieta de la Astronom\'ia s/n, 18008, Granada, Spain}
\author{Lucio A. Antonelli}
\affiliation{INAF Osservatorio Astronomico di Roma, Via Frascati 33, 00078 Monte Porzio Catone (RM), Italy}
\affiliation{Space Science Data Center, Agenzia Spaziale Italiana, Via del Politecnico snc, 00133 Roma, Italy}
\author{Matteo Bachetti}
\affiliation{INAF Osservatorio Astronomico di Cagliari, Via della Scienza 5, 09047 Selargius (CA), Italy}
\author{Luca Baldini}
\affiliation{Istituto Nazionale di Fisica Nucleare, Sezione di Pisa, Largo B. Pontecorvo 3, 56127 Pisa, Italy}
\affiliation{Dipartimento di Fisica, Universit\`{a} di Pisa, Largo B. Pontecorvo 3, 56127 Pisa, Italy}
\author{Wayne H. Baumgartner}
\affiliation{NASA Marshall Space Flight Center, Huntsville, AL 35812, USA}
\author{Ronaldo Bellazzini}
\affiliation{Istituto Nazionale di Fisica Nucleare, Sezione di Pisa, Largo B. Pontecorvo 3, 56127 Pisa, Italy}
\author{Stephen D. Bongiorno}
\affiliation{NASA Marshall Space Flight Center, Huntsville, AL 35812, USA}
\author{Raffaella Bonino}
\affiliation{Istituto Nazionale di Fisica Nucleare, Sezione di Torino, Via Pietro Giuria 1, 10125 Torino, Italy}
\affiliation{Dipartimento di Fisica, Universit\`{a} degli Studi di Torino, Via Pietro Giuria 1, 10125 Torino, Italy}
\author{Alessandro Brez}
\affiliation{Istituto Nazionale di Fisica Nucleare, Sezione di Pisa, Largo B. Pontecorvo 3, 56127 Pisa, Italy}
\author{Niccolò Bucciantini}
\affiliation{INAF Osservatorio Astrofisico di Arcetri, Largo Enrico Fermi 5, 50125 Firenze, Italy}
\affiliation{Dipartimento di Fisica e Astronomia,  Universit\`{a} degli Studi di Firenze, Via Sansone 1, 50019 Sesto Fiorentino (FI), Italy}
\affiliation{Istituto Nazionale di Fisica Nucleare, Sezione di Firenze, Via Sansone 1, 50019 Sesto Fiorentino (FI), Italy}
\author{Fiamma Capitanio}
\affiliation{INAF Istituto di Astrofisica e Planetologia Spaziali, Via del Fosso del Cavaliere 100, 00133 Roma, Italy}
\author{Simone Castellano}
\affiliation{Istituto Nazionale di Fisica Nucleare, Sezione di Pisa, Largo B. Pontecorvo 3, 56127 Pisa, Italy}
\author{Elisabetta Cavazzuti}
\affiliation{ASI - Agenzia Spaziale Italiana, Via del Politecnico snc, 00133 Roma, Italy}
\author{Stefano Ciprini}
\affiliation{Istituto Nazionale di Fisica Nucleare, Sezione di Roma "Tor Vergata", Via della Ricerca Scientifica 1, 00133 Roma, Italy}
\affiliation{Space Science Data Center, Agenzia Spaziale Italiana, Via del Politecnico snc, 00133 Roma, Italy}
\author{Enrico Costa}
\affiliation{INAF Istituto di Astrofisica e Planetologia Spaziali, Via del Fosso del Cavaliere 100, 00133 Roma, Italy}
\author{Alessandra De Rosa}
\affiliation{INAF Istituto di Astrofisica e Planetologia Spaziali, Via del Fosso del Cavaliere 100, 00133 Roma, Italy}
\author{Ettore Del Monte}
\affiliation{INAF Istituto di Astrofisica e Planetologia Spaziali, Via del Fosso del Cavaliere 100, 00133 Roma, Italy}
\author{Laura Di Gesu}
\affiliation{ASI - Agenzia Spaziale Italiana, Via del Politecnico snc, 00133 Roma, Italy}
\author{Niccolò Di Lalla}
\affiliation{Department of Physics and Kavli Institute for Particle Astrophysics and Cosmology, Stanford University, Stanford, California 94305, USA}
\author{Alessandro Di Marco}
\affiliation{INAF Istituto di Astrofisica e Planetologia Spaziali, Via del Fosso del Cavaliere 100, 00133 Roma, Italy}
\author{Immacolata Donnarumma}
\affiliation{ASI - Agenzia Spaziale Italiana, Via del Politecnico snc, 00133 Roma, Italy}
\author{Victor Doroshenko}
\affiliation{Institut f\"{u}r Astronomie und Astrophysik, Universit\"{a}t T\"{u}bingen, Sand 1, 72076 T\"{u}bingen, Germany}
\affiliation{Space Research Institute of the Russian Academy of Sciences, Profsoyuznaya Str. 84/32, Moscow 117997, Russia}
\author{Michal Dovčiak}
\affiliation{Astronomical Institute of the Czech Academy of Sciences, Boční II 1401/1, 14100 Praha 4, Czech Republic}
\author{Teruaki Enoto}
\affiliation{RIKEN Cluster for Pioneering Research, 2-1 Hirosawa, Wako, Saitama 351-0198, Japan}
\author{Yuri Evangelista}
\affiliation{INAF Istituto di Astrofisica e Planetologia Spaziali, Via del Fosso del Cavaliere 100, 00133 Roma, Italy}
\author{Sergio Fabiani}
\affiliation{INAF Istituto di Astrofisica e Planetologia Spaziali, Via del Fosso del Cavaliere 100, 00133 Roma, Italy}
\author{Javier A. Garcia}
\affiliation{California Institute of Technology, Pasadena, CA 91125, USA}
\author{Shuichi Gunji}
\affiliation{Yamagata University,1-4-12 Kojirakawa-machi, Yamagata-shi 990-8560, Japan}
\author{Kiyoshi Hayashida}
\affiliation{Osaka University, 1-1 Yamadaoka, Suita, Osaka 565-0871, Japan}
\author{Jeremy Heyl}
\affiliation{University of British Columbia, Vancouver, BC V6T 1Z4, Canada}
\author{Wataru Iwakiri}
\affiliation{Department of Physics, Faculty of Science and Engineering, Chuo University, 1-13-27 Kasuga, Bunkyo-ku, Tokyo 112-8551, Japan}
\author{Svetlana G. Jorstad}
\affiliation{Institute for Astrophysical Research, Boston University, 725 Commonwealth Avenue, Boston, MA 02215, USA}
\affiliation{Department of Astrophysics, St. Petersburg State University, Universitetsky pr. 28, Petrodvoretz, 198504 St. Petersburg, Russia}
\author{Vladimir Karas}
\affiliation{Astronomical Institute of the Czech Academy of Sciences, Boční II 1401/1, 14100 Praha 4, Czech Republic}
\author{Takao Kitaguchi}
\affiliation{RIKEN Cluster for Pioneering Research, 2-1 Hirosawa, Wako, Saitama 351-0198, Japan}
\author{Jeffery J. Kolodziejczak}
\affiliation{NASA Marshall Space Flight Center, Huntsville, AL 35812, USA}
\author{Henric Krawczynski}
\affiliation{Physics Department and McDonnell Center for the Space Sciences, Washington University in St. Louis, St. Louis, MO 63130, USA}
\author{Fabio La Monaca}
\affiliation{INAF Istituto di Astrofisica e Planetologia Spaziali, Via del Fosso del Cavaliere 100, 00133 Roma, Italy}
\author{Luca Latronico}
\affiliation{Istituto Nazionale di Fisica Nucleare, Sezione di Torino, Via Pietro Giuria 1, 10125 Torino, Italy}
\author{Simone Maldera}
\affiliation{Istituto Nazionale di Fisica Nucleare, Sezione di Torino, Via Pietro Giuria 1, 10125 Torino, Italy}
\author{Alberto Manfreda}
\affiliation{Istituto Nazionale di Fisica Nucleare, Sezione di Pisa, Largo B. Pontecorvo 3, 56127 Pisa, Italy}
\author{Francesco Massaro}
\affiliation{Istituto Nazionale di Fisica Nucleare, Sezione di Torino, Via Pietro Giuria 1, 10125 Torino, Italy}
\affiliation{Dipartimento di Fisica, Università degli Studi di Torino, Via Pietro Giuria 1, 10125 Torino, Italy}
\author{Ikuyuki Mitsuishi}
\affiliation{Graduate School of Science, Division of Particle and Astrophysical Science, Nagoya University, Furo-cho, Chikusa-ku, Nagoya, Aichi 464-8602, Japan}
\author{Tsunefumi Mizuno}
\affiliation{Hiroshima Astrophysical Science Center, Hiroshima University, 1-3-1 Kagamiyama, Higashi-Hiroshima, Hiroshima 739-8526, Japan}
\author{Fabio Muleri}
\affiliation{INAF Istituto di Astrofisica e Planetologia Spaziali, Via del Fosso del Cavaliere 100, 00133 Roma, Italy}
\author{Michela Negro}
\affiliation{University of Maryland, Baltimore County, Baltimore, MD 21250, USA}
\affiliation{NASA Goddard Space Flight Center, Greenbelt, MD 20771, USA}
\affiliation{Center for Research and Exploration in Space Science and Technology, NASA/GSFC, Greenbelt, MD 20771, USA}
\author{C.-Y. Ng}
\affiliation{Department of Physics, The University of Hong Kong, Pokfulam, Hong Kong}
\author{Stephen L. O'Dell}
\affiliation{NASA Marshall Space Flight Center, Huntsville, AL 35812, USA}
\author{Nicola Omodei}
\affiliation{Department of Physics and Kavli Institute for Particle Astrophysics and Cosmology, Stanford University, Stanford, California 94305, USA}
\author{Chiara Oppedisano}
\affiliation{Istituto Nazionale di Fisica Nucleare, Sezione di Torino, Via Pietro Giuria 1, 10125 Torino, Italy}
\author{Alessandro Papitto}
\affiliation{INAF Osservatorio Astronomico di Roma, Via Frascati 33, 00078 Monte Porzio Catone (RM), Italy}
\author{George G. Pavlov}
\affiliation{Department of Astronomy and Astrophysics, Pennsylvania State University, University Park, PA 16802, USA}
\author{Abel L. Peirson}
\affiliation{Department of Physics and Kavli Institute for Particle Astrophysics and Cosmology, Stanford University, Stanford, California 94305, USA}
\author{Melissa Pesce-Rollins}
\affiliation{Istituto Nazionale di Fisica Nucleare, Sezione di Pisa, Largo B. Pontecorvo 3, 56127 Pisa, Italy}
\author{Maura Pilia}
\affiliation{INAF Osservatorio Astronomico di Cagliari, Via della Scienza 5, 09047 Selargius (CA), Italy}
\author{Andrea Possenti}
\affiliation{INAF Osservatorio Astronomico di Cagliari, Via della Scienza 5, 09047 Selargius (CA), Italy}
\author{Brian D. Ramsey}
\affiliation{NASA Marshall Space Flight Center, Huntsville, AL 35812, USA}
\author{John Rankin}
\affiliation{INAF Istituto di Astrofisica e Planetologia Spaziali, Via del Fosso del Cavaliere 100, 00133 Roma, Italy}
\author{Ajay Ratheesh}
\affiliation{INAF Istituto di Astrofisica e Planetologia Spaziali, Via del Fosso del Cavaliere 100, 00133 Roma, Italy}
\author{Roger W. Romani}
\affiliation{Department of Physics and Kavli Institute for Particle Astrophysics and Cosmology, Stanford University, Stanford, California 94305, USA}
\author{Carmelo Sgrò}
\affiliation{Istituto Nazionale di Fisica Nucleare, Sezione di Pisa, Largo B. Pontecorvo 3, 56127 Pisa, Italy}
\author{Patrick Slane}
\affiliation{Center for Astrophysics, Harvard \& Smithsonian, 60 Garden St, Cambridge, MA 02138, USA}
\author{Paolo Soffitta}
\affiliation{INAF Istituto di Astrofisica e Planetologia Spaziali, Via del Fosso del Cavaliere 100, 00133 Roma, Italy}
\author{Gloria Spandre}
\affiliation{Istituto Nazionale di Fisica Nucleare, Sezione di Pisa, Largo B. Pontecorvo 3, 56127 Pisa, Italy}
\author{Toru Tamagawa}
\affiliation{RIKEN Cluster for Pioneering Research, 2-1 Hirosawa, Wako, Saitama 351-0198, Japan}
\author{Fabrizio Tavecchio}
\affiliation{INAF Osservatorio Astronomico di Brera, Via E. Bianchi 46, 23807 Merate (LC), Italy}
\author{Roberto Taverna}
\affiliation{Dipartimento di Fisica e Astronomia,Universit\`{a} degli Studi di Padova, Via Marzolo 8, 35131 Padova, Italy}
\author{Yuzuru Tawara}
\affiliation{Graduate School of Science, Division of Particle and Astrophysical Science, Nagoya University, Furo-cho, Chikusa-ku, Nagoya, Aichi 464-8602, Japan}
\author{Allyn F. Tennant}
\affiliation{NASA Marshall Space Flight Center, Huntsville, AL 35812, USA}
\author{Nicholas E. Thomas}
\affiliation{NASA Marshall Space Flight Center, Huntsville, AL 35812, USA}
\author{Francesco Tombesi}
\affiliation{Dipartimento di Fisica,Universit\`{a} egli Studi di Roma "Tor Vergata", Via della Ricerca Scientifica 1, 00133 Roma, Italy}
\affiliation{Istituto Nazionale di Fisica Nucleare, Sezione di Roma "Tor Vergata", Via della Ricerca Scientifica 1, 00133 Roma, Italy}
\affiliation{Department of Astronomy, University of Maryland, College Park, Maryland 20742, USA}
\author{Alessio Trois}
\affiliation{INAF Osservatorio Astronomico di Cagliari, Via della Scienza 5, 09047 Selargius (CA), Italy}
\author{Sergey Tsygankov}
\affiliation{Department of Physics and Astronomy, 20014 University of Turku, Finland}
\affiliation{Space Research Institute of the Russian Academy of Sciences, Profsoyuznaya Str. 84/32, Moscow 117997, Russia}
\author{Roberto Turolla}
\affiliation{Dipartimento di Fisica e Astronomia,Universit\`{a} degli Studi di Padova, Via Marzolo 8, 35131 Padova, Italy}
\affiliation{Mullard Space Science Laboratory, University College London, Holmbury St Mary, Dorking, Surrey RH5 6NT, UK}
\author{Jacco Vink}
\affiliation{Anton Pannekoek Institute for Astronomy \& GRAPPA, University of Amsterdam, Science Park 904, 1098 XH Amsterdam, The Netherlands}
\author{Martin C. Weisskopf}
\affiliation{NASA Marshall Space Flight Center, Huntsville, AL 35812, USA}
\author{Fei Xie}
\affiliation{Guangxi Key Laboratory for Relativistic Astrophysics, School of Physical Science and Technology, Guangxi University, Nanning 530004, China}
\author{Silvia Zane}
\affiliation{Mullard Space Science Laboratory, University College London, Holmbury St Mary, Dorking, Surrey RH5 6NT, UK}
\collaboration{18}{(IXPE Collaboration)}
\nocollaboration{3}
\author{James Rodi}
\affiliation{INAF Istituto di Astrofisica e Planetologia Spaziali, Via del Fosso del Cavaliere 100, 00133 Roma, Italy}
\author{Elisabeth Jourdain}
\affiliation{CNRS, Institut de Recherche en Astrophysique et Planétologie (IRAP), 9 Av. colonel Roche, BP 44346, F-31028 Toulouse cedex 4, France}
\author{Jean-Pierre Roques}
\affiliation{CNRS, Institut de Recherche en Astrophysique et Planétologie (IRAP), 9 Av. colonel Roche, BP 44346, F-31028 Toulouse cedex 4, France}

%% Note that the \and command from previous versions of AASTeX is now
%% depreciated in this version as it is no longer necessary. AASTeX 
%% automatically takes care of all commas and "and"s between authors names.

%% AASTeX 6.31 has the new \collaboration and \nocollaboration commands to
%% provide the collaboration status of a group of authors. These commands 
%% can be used either before or after the list of corresponding authors. The
%% argument for \collaboration is the collaboration identifier. Authors are
%% encouraged to surround collaboration identifiers with ()s. The 
%% \nocollaboration command takes no argument and exists to indicate that
%% the nearby authors are not part of surrounding collaborations.

%% Mark off the abstract in the ``abstract'' environment. 
\begin{abstract}

We present measurements of the polarization of X-rays in the $2–8 \keV$ band from the nucleus of the radio galaxy Centaurus A (Cen A), using a 100~ks observation from the Imaging X-ray Polarimetry Explorer (IXPE ). Nearly simultaneous observations of Cen A were also taken with the \textit{Swift}, \textit{NuSTAR}, and \textit{INTEGRAL} observatories. No statistically significant degree of polarization is detected with IXPE. These observations have a minimum detectable polarization at $99\%$ confidence (MDP$_{99}$) of $6.5\%$ using a weighted, spectral model-independent calculation in the $2–8 \keV$ band. The polarization angle $\psi$ is consequently unconstrained. Spectral fitting across three orders of magnitude in X-ray energy ($0.3–400 \keV$) demonstrates that the SED of Cen A is well described by a simple power law with moderate intrinsic absorption ($N_\mathrm{H} \sim 10^{23} \cm^{-2}$) and a Fe K$\alpha$ emission line, although a second unabsorbed power law is required to account for the observed spectrum at energies below $2 \keV$. This spectrum suggests that the reprocessing material responsible for this emission line is optically thin and distant from the central black hole. Our upper limits on the X-ray polarization are consistent with the predictions of Compton scattering, although the specific seed photon population responsible for production of the X-rays cannot be identified. The low polarization degree, variability in the core emission, and the relative lack of variability in the Fe K$\alpha$ emission line support a picture where electrons are accelerated in a region of highly disordered magnetic fields surrounding the innermost jet.

%and the geometry of the radio jets on the smallest scales determined by %probed by 
%the Event Horizon Telescope. Fixing the polarization angle to a position angle parallel/perpendicular to the small scale jet gives a polarization degree of $AA \%$. Such an agreement/disagreement between the X-ray polarization and radio/IR polarization measurements, and jet position angle suggests that the non-thermal X-ray emission originates primarily in the accretion disk/jets of the galaxy. 

\end{abstract}

%% Keywords should appear after the \end{abstract} command. 
%% The AAS Journals now uses Unified Astronomy Thesaurus concepts:
%% https://astrothesaurus.org
%% You will be asked to selected these concepts during the submission process
%% but this old "keyword" functionality is maintained in case authors want
%% to include these concepts in their preprints.

%% From the front matter, we move on to the body of the paper.
%% Sections are demarcated by \section and \subsection, respectively.
%% Observe the use of the LaTeX \label
%% command after the \subsection to give a symbolic KEY to the
%% subsection for cross-referencing in a \ref command.
%% You can use LaTeX's \ref and \label commands to keep track of
%% cross-references to sections, equations, tables, and figures.
%% That way, if you change the order of any elements, LaTeX will
%% automatically renumber them.
%%
%% We recommend that authors also use the natbib \citep
%% and \citet commands to identify citations.  The citations are
%% tied to the reference list via symbolic KEYs. The KEY corresponds
%% to the KEY in the \bibitem in the reference list below. 

\section{Introduction} \label{sec:intro}

Particle acceleration is 
ubiquitous 
%a ubiquitous process 
in astrophysical systems, with charged particles routinely being accelerated to energies orders of magnitude larger than those attainable in terrestrial experiments. Supermassive black holes ($M \gtrsim 10^{6} M_{\bigodot}$) at the centers of galaxies 
are known 
%are expected 
to be among the most prominent 
particle acceleration sites in the universe,  
%sites of particle acceleration, 
%although many 
although details of 
%the underlying processes 
the acceleration and radiative mechanisms 
in active galactic nuclei (AGN) remain ambiguous: multiple models of these processes can fit the spectral energy distributions (SEDs) of AGN equally well. Polarimetry, which more directly constrains the magnetic field geometry of the regions where acceleration 
%is taking 
takes place, 
is able to break many of these degeneracies. 
Polarization measurements of high-energy, non-thermal emission 
%impose crucial constraints on  
provide an additional means to determine 
the radiative processes 
and the nature of the populations of energetic particles 
that produce them.

Centaurus A is the most nearby radio galaxy ($z = 0.0018, \thinspace D= 3.84 \Mpc$), and the radio jets in its nucleus have been studied in great detail. Radio observations of the core of Cen A indicate complex velocity and polarization structures. Long-term observations of Cen A with Very Long Baseline Interferometry \citep[VLBI,][]{Tingay2001, Mueller2014} suggest that within the central $\sim 1 \pc$ the jets have an apparent velocity (normalized to the speed of light) of $\beta \sim 0.12$, with a jet viewing angle of $\sim 12^{\circ} - 45^{\circ}$ with respect to the line of sight. On larger scales of $\sim 100 \pc$, \cite{Hardcastle2003} measured apparent velocities of $\beta \sim 0.5$. Cen A has also recently been resolved on an angular scale of $25 \mu$arcsec at a wavelength of 1.3 mm with the Event Horizon Telescope \citep[EHT,][]{Janssen2021}. These observations reveal both an edge-brightened jet northeast of the core and a fainter counter-jet  to the southwest. The surface brightness structure of these jets is interpreted in \cite{Janssen2021} as evidence for a helical magnetic field wrapped around the jets.  The jet and counter-jet each have two branches that are clearly separated from one another. For the jet these two branches have position angles of $43^{\circ}$ and $52^{\circ}$ (measured eastward from north), which is nearly identical to the jet position angle on larger scales \citep[55$^\circ$ $\pm$ 7$^\circ$; e.g. ][]{Schreier1979,Burns1983,Jones1996}. X-ray observations of the nucleus of Cen A with \textit{Chandra} and \textit{XMM-Newton} show that the X-rays are mainly generated within a small region around the nucleus, as the core has previously been measured to vary in luminosity by $\sim 60\%$ on time scales of $\sim 20$ days \citep{Evans2004}. A distance of 20 light-days ($\sim 0.0168 \pc$) corresponds to an angular scale of $\sim 0.001^{\prime \prime}$ at the distance of Cen A, far smaller than the best angular resolution provided by \textit{Chandra} ($\sim 0.5^{\prime \prime}$) for X-ray imaging, but well matched to the resolution of the EHT radio observations. The position angle of the inner radio jet can therefore be compared with the electric-vector position angle (EVPA, itself perpendicular to the local magnetic field) of Cen A. 

Measurements of radio polarization by \cite{Burns1983} at $6 \cm$ show that the polarization degree varies between  $\sim 5-20\%$ in different regions in the vicinity of the nucleus, although the nucleus itself is only polarized at the $0.21 \pm 0.01\%$ level. The core region of Cen A was measured to have a polarization of $0.46 \pm 0.03\%$ at $20 \cm$ in the same study. The morphology of the polarization vectors throughout the larger scale ($\sim 100 \pc$) jet regions suggests that the magnetic field lines are consistently parallel to the jet axis, with no evidence of field lines transverse to the jet near its outer edges \citep{Clarke1986, Hardcastle2003}. More recent ALMA results find that the core of Cen A has no measurable polarization at 221 GHz \citep{Goddi2021}, with a $3-\sigma$ upper limit to the linear polarization of $0.09\% $. Far-infrared polarimetry with SOFIA/HAWC+ (89 $\mu$m) measured a polarization of $1.5\pm$ 0.2\% with an EVPA of $151\pm4^{\circ}$, consistent with dichroic absorption and emission \citep{Lopez2021}. After correction for the stellar component of the host galaxy, the $\sim1''$ core of Cen A was previously observed to have a polarization degree of $\sim11.1\pm0.2\% $ at a wavelength of  $\sim2 \thinspace \mu$m \citep[][]{Capetti2000}, with an EVPA of $148^{\circ}$ east of north, which is perpendicular to the jet position angle. 

Past observations have shown that the X-ray spectrum is affected by absorption below $\sim 10 \keV$, with a column density of $N_{\rm H}\sim  10^{23} \cm^{-2}$. At hard X-rays the continuum photon flux is characterized by a power law with slope $\Gamma \sim 1.8$ \citep[e.g.][and references therein]{furst16}. The presence of a narrow Fe K emission line, combined with the absence of a strong Compton hump, suggests reflection from Compton-thin material.  The detection of a high energy cut-off or continuum curvature, from hard X-ray satellites such as \textit{CGRO}/OSSE, \textit{Suzaku}, \textit{INTEGRAL}, and \textit{NuSTAR} \citep{Kinzer95,steinle98,markowitz07,beckmann11,furst16}, is under debate, which prevents an unambiguous determination of the origin of the primary X-ray emission. It could be dominated either by thermal Compton emission from a hot X-ray ``corona'', as in radio-quiet AGN \citep{1995ApJ...449L..13S,1997ESASP.382..373Z,2000ApJ...542..703Z}, or by non-thermal emission from the base of the jet, or a combination of the two.

Here we report polarimetric observations of Cen A obtained with the \textit{Imaging X-ray Polarimetry Explorer} (\textit{IXPE}), a NASA mission in partnership with the Italian Space Agency (ASI). As described in detail elsewhere (\cite{Weisskopf2022} and references therein), \textit{IXPE} includes three identical X-ray telescopes, each comprising an X-ray mirror assembly and a polarization-sensitive pixelated detector \citep{Soffitta2021}, to provide imaging polarimetry over a nominal $2-8 \keV$ band. \textit{IXPE} data telemetered to ground stations in Malindi (primary) and in Singapore (secondary) are transmitted to the Mission Operations Center (MOC, at the Laboratory for Atmospheric and Space Physics, University of Colorado) and then to the Science Operations Center (SOC, at NASA Marshall Space Flight Center). The SOC processes science and relevant engineering and ancillary data, to produce data products that are archived at the High-Energy Astrophysics Science Archive Research Center (HEASARC, at NASA Goddard Space Flight Center), for use by the international astrophysics community. Cen A was observed for a net exposure time of $100 \ks$, from 2022 February 15 (01:00 UT) to 17 (13:00 UT).

In addition to the \textit{IXPE} observations, Cen A was also observed simultaneously (or nearly simultaneously) by the \textit{NuSTAR}, \textit{Swift}, and \textit{INTEGRAL} missions in order to provide broad-band ($\sim$0.3--400~keV) measurements of the X-ray spectrum.

\section{Observations}

\textit{IXPE} observations of Cen A were taken starting at 00:00 UTC on 2022 February 15. Simultaneous (or nearly so) observations with other X-ray telescopes are summarized in Table \ref{tab:observations}. Operational constraints limited the ability of \textit{NuSTAR} and \textit{Swift} to observe Cen A simultaneously with \textit{IXPE}. Given that Cen A has not been previously observed to exhibit substantial variability in the X-rays on timescales of days, we do not anticipate this offset in time to introduce any major systematic errors in our joint spectral analysis. The light curve during each of these observations also show no evidence for significant intra-observation variability in the emission from Cen A. 

\begin{table}[htbp]
\centering  
\begin{tabular}{cccccccc}
\hline \hline
Mission	   & Obs ID & Energy  & Start date    & End date      & Exposure 	 \\
	&	  & $\keV$  & (YYYY-MM-DD)  & (YYYY-MM-DD)  & (ks)  	\\
\hline
\textit{IXPE}	& 01004301 & $2-8 $	 	        & 2022-02-15    & 2022-02-17    & $100$ 	\\
\textit{NuSTAR} & 60701033002 & $ 5 - 70 $ & 2022-02-17 & 2022-02-17 & $20.8$  \\
\textit{NuSTAR} & 10802601002 & $ 5 - 70 $ & 2022-02-18 & 2022-02-18 & $23.1$  \\
\textit{Swift} & 00096451017 & $0.3 - 8 $ & 2022-02-24 & 2022-02-25 & $ 0.59$ \\
\textit{Swift} & 00096451018 & $0.3 - 8 $ & 2022-02-25 & 2022-02-25 &  $1.4$ \\
\textit{INTEGRAL}/ISGRI & & $30 - 400 $ & 2022-02-11 & 2022-02-18 & $ 144$ \\
\textit{INTEGRAL}/SPI & & $20 - 400 $ & 2022-01-11 & 2022-02-18 & $ 672$ \\
\hline \hline
\end{tabular}
\caption{Observational log of the X-ray telescope observations used in this work. }     
\label{tab:observations}   
\end{table}

\subsection{IXPE}
\label{sec:ixpe}

We refer interested readers to \cite{Weisskopf2016, Weisskopf2022} for a complete description of the hardware deployed upon \textit{IXPE} and its performance, and here only summarize the most relevant details for this work. \textit{IXPE} data were processed using a pipeline that applies estimates of the photoelectron emission direction (and hence the polarization), location, and energy of each event after applying corrections for charging effects, detector temperature, and Gas Electron Multiplier (GEM) gain non-uniformity. Spurious modulation is removed using the algorithm of \cite{Rankin2022}. 

 The output of this pipeline processing is an event file for each of the three \textit{IXPE} Detector Units (DUs) that contains, in addition to the typical information related to spatially resolved X-ray astronomy, the event-by-event Stokes parameters, from which the polarization of the radiation can be derived. In addition to the standard pipeline processing for \textit{IXPE} data, three additional corrections were applied to the data: small time-dependent changes to the gain correction, obtained from data taken with the on-board calibration sources \citep{Ferrazzoli2020} close to the actual time of observation; further corrections to the aspect solution to remove apparent positional shifts in sources arising from small motions of the \textit{IXPE} boom; and the weights of each event, updated to reflect the results of \cite{DiMarco2022} and to fix a known error in the data processing pipeline at the time of processing. 
 
 Analysis of the processed \textit{IXPE} event lists utilizing both weighted and unweighted data was carried out with multiple independent analysis tools, most notably \texttt{ixpeobssim} \citep{Baldini2022}. The \texttt{ixpeobssim} software suite was designed specifically to operate with both simulated and real \textit{IXPE} data; \cite{Baldini2022} describe these algorithms in detail. For Cen A, the core region was filtered from the rest of the event list by using a $1^{\prime}$ radius aperture\footnote{This radius corresponds to $\sim 2$ half-power diameters for the \textit{IXPE} optics.} around the centroid of the X-ray data.  
 
 We have attempted to measure the polarization degree and angle and their uncertainties along with the minimum detectable polarization at 99\% confidence ($\mathrm{MDP}_{99}$) of the region using several different techniques in order to test the robustness of the final result. The simplest calculation was performed with a polarization cube, which is constructed using the \textit{xpbin} tool on the selected event list with the PCUBE algorithm. The polarization properties and their uncertainties were calculated according to the equations contained in \cite{Kislat2015}.  No treatment of the expected background counts in the source region was performed for this calculation. Estimates derived from multiple source-free regions of the sky using an identical $1'$ radius aperture suggest that background is $< 10\%$ of the total counts within the source aperture. We have constructed two polarization cubes from these observations - one that utilizes the event-specific weights and one that does not. 
 
 Spectro-polarimetric fitting was also used to determine the polarization degree and angle of the Cen A core. Spectra were extracted using the \textit{xpbin} tool on the same event list by using the PHA algorithm to create OGIP-compliant spectra for the total (Stokes I) and Stokes $Q/U$ parameters. For the spectral extraction, a local source-free circular aperture with a radius of $1'$ was used to construct background Stokes spectra. Two variants of the 9 \textit{IXPE} spectra (3 detectors $\times$ $I/Q/U$ spectra per detector) were produced: one variant that includes the event weights and a second that does not. In this paper all references to spectro-polarimetric fitting  will be based on the unweighted spectra. Tests using the weighted spectra have confirmed that both sets of spectra result in upper limits to the polarization, with no meaningful differences between them. We have also confirmed that the best-fit spectral model parameters (including polarization) are not sensitive to the particular choice of background region.

%% Note to Steven: I added two bib entries to sample631.bib that are needed for this paragraph.  The main bib file seems to be locked against editing.
As a further independent check of the polarization measurements, we also utilized a maximum-likelihood calculation of the polarization degree and angle based on the method developed by \citet{marshall20} to allow for background selected from a separate region (Marshall, in prep). Background events are accounted for in a manner similar to the calculations in \cite{Elsner2012}, although this new calculation explicitly assumes Poisson rather than Gaussian statistics as assumed in \cite{Elsner2012} . These calculations were developed and executed independently of \texttt{ixpeobssim}.  The unweighted energy-dependent modulation factor $\mu(E)$ from \citet{DiMarco2022} was used and no events were excluded from the calculation.  For this calculation, background was estimated from an annular region 160-320\arcsec\ from the core.

\subsection{NuSTAR}

To calibrate and clean the \textit{NuSTAR} data, we used the NuSTAR Data Analysis Software (\texttt{NuSTARDAS}, v. 2.1.1). We derived level 2 cleaned event files with the \texttt{nupipeline} task, while high-level scientific data products were extracted using the \texttt{nuproducts} tool. The latest calibration data files in the \textit{NuSTAR} CALDB (version 20220301) were used.  These data have a significantly lower count rate than the data analyzed in \cite{furst16}.  Our measured count rate for Cen A in the $3-70 \keV$ energy band is $\sim 4 \thinspace \mathrm{ct} \s^{-1}$, which is a factor of $\sim 4$ lower than the value quoted for the 2013 \textit{NuSTAR} observation \citep{furst16}. A $70''$ radius circular region (corresponding to a 90\% enclosed energy fraction of the \textit{NuSTAR} point-spread function) was used to extract the source spectrum\footnote{We note that this source aperture is slightly smaller than the $100''$ radius used for spectral extraction in \cite{furst16}, but this discrepancy alone cannot account for the lower count rate we measure.}. A concentric annulus with inner and outer radii of $210''$ and $270''$ was then adopted to estimate the background, which subtends a solid angle a factor of 2 larger than the $120''$ radius circle  used for the background region in \cite{furst16}.  FPMA/B spectra were then binned to achieve a S/N ratio larger than 10 in each spectral channel.

\subsection{Swift}

\textit{Swift}-XRT observations were performed in Windowed Timing (WT) mode, and the data were processed using the X-Ray Telescope Data Analysis Software (\texttt{XRTDAS}, v. 3.6.1). As for \textit{NuSTAR} data, we used the latest calibration files available in the \textit{Swift}/XRT CALDB (version 20210915). The X-ray source spectrum was extracted from the cleaned event file using a circular region with a radius of 20-pixels ($\sim$ 46 arcsec). This region corresponds to a 90\% enclosed energy fraction of the \textit{Swift} point-spread function. The background was extracted using a circular region of 20-pixel radius from a blank sky WT observation available in the \textit{Swift} archive. Finally, the 0.3--10 keV source spectra were binned requiring each spectral channel to have at least 25 counts.

\subsection{INTEGRAL}

The \textit{INTEGRAL} observations used in these analyses are part of a multi-epoch program. The data reported here consists of exposures that followed a \( 5 \times 5\) dithering pattern.  Data contaminated by solar flares and pointings affected by Earth's radiation belt have been removed.

The SPI data from 2022 January 11 to February 18 have been processed through the SPIDAI interface\footnote{Interface developed at IRAP to analyze SPI data, publicly vailable at http://sigma-2.cesr.fr/integral/spidai; see description in \citet{burke2014}}.  Cen A was assumed to be the only source in the sky model, and constant during the exposure duration.  Due to the evolution of the background over the entire data set, its normalization was determined on a \(\sim 1\) hr timescale.  The spectrum was constructed in 21 channels from 20 to 500 keV and averaged over the exposure. Tests comparing the full SPI data set to the SPI spectrum extracted for a time window immediately before and after \textit{IXPE} observations show no significant differences in count rate for any channel. We therefore choose to utilize data taken throughout this entire time window to maximize the spectrum's signal-to-noise, especially for the highest energy  channels.

The ISGRI data between 2022 February 11-18 were processed using the Off-line Science Analysis 11.2 version (OSA11.2), resulting in a total useful duration of 144.4 ks.  The source was considered constant on a $\sim1800$ s timescale.  The spectrum was averaged over the observing period and binned into 29 channels in the 30-1000 keV energy range.  

\section{Results}
\begin{figure}
\centering
\includegraphics[width=0.65\textwidth]{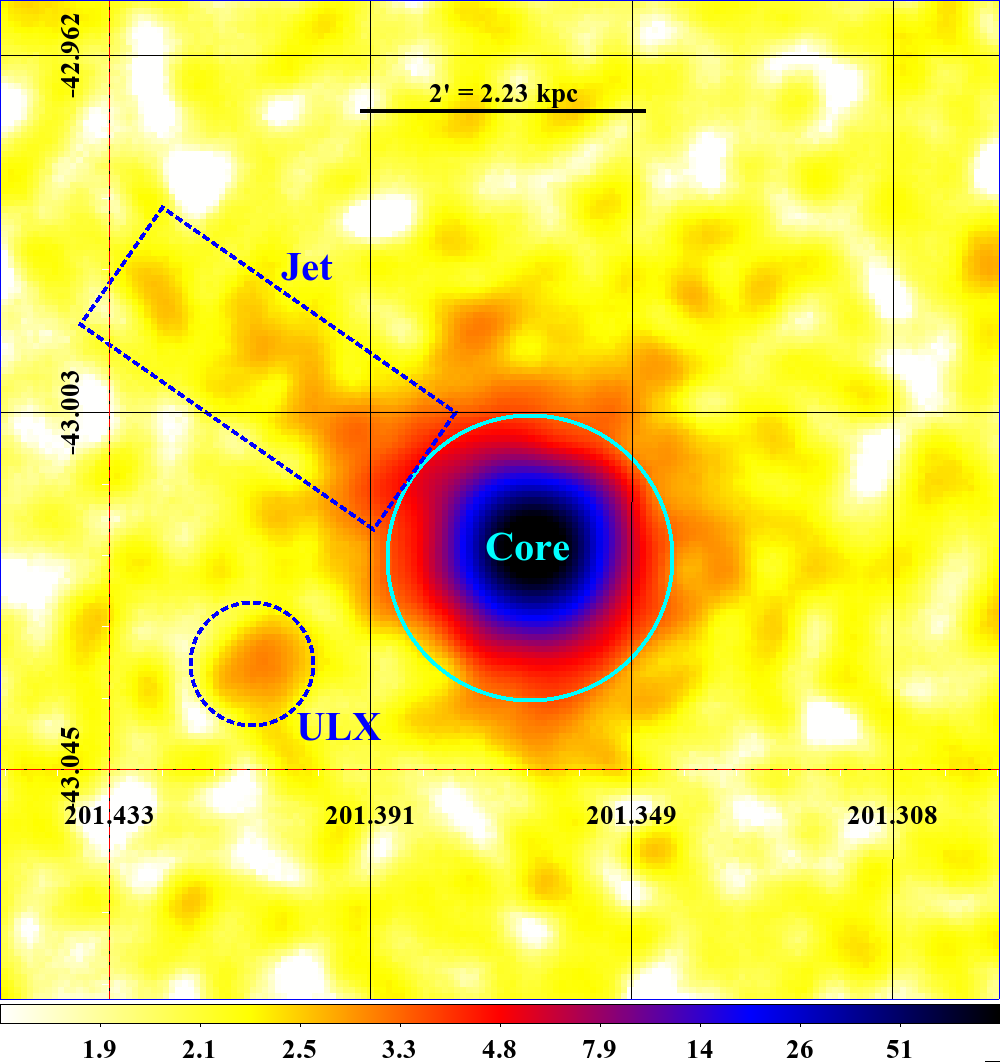}
\caption{Image of Cen A constructed using the data from all three IXPE detectors. The image has been smoothed using a Gaussian  with a width of $\sigma = 3$ pixels and a radius of 6 pixels. The regions used to calculate the MDP$_{99}$ of the jet and ULX regins are shown in a blue dashed box and circle, respectively. The cyan region is the $1^{\prime}$ aperture surrounding the core used for investigating the polarization of the core. }
\label{fig::CenAImage}
\end{figure}

Figure \ref{fig::CenAImage} displays the combined image of Cen A resulting from data obtained with all three \textit{IXPE} detectors. Three regions of interest are shown on this figure: the core, the X-ray jet, and a previously identified ultra-luminous X-ray source (ULX) to the southeast of the core. Between all three detectors, a total of 28179 (source + background) counts were detected in the core region, 2147 in the jet, and 469 in the vicinity of the ULX.  This work will focus on the core region but briefly discuss upper limits to X-ray polarization in the other two regions as well. 

\subsection{Polarization}

\begin{figure}
\centering
\includegraphics[width=0.48\textwidth]{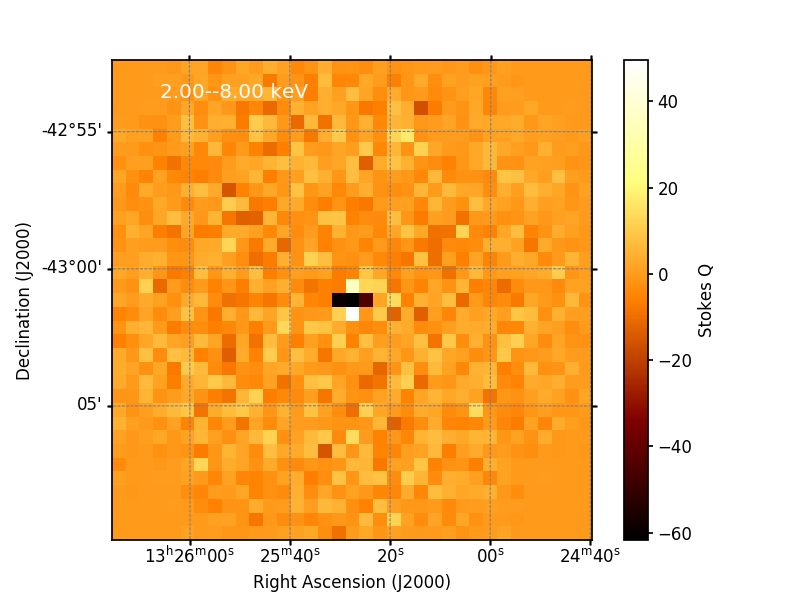}
\includegraphics[width=0.48\textwidth]{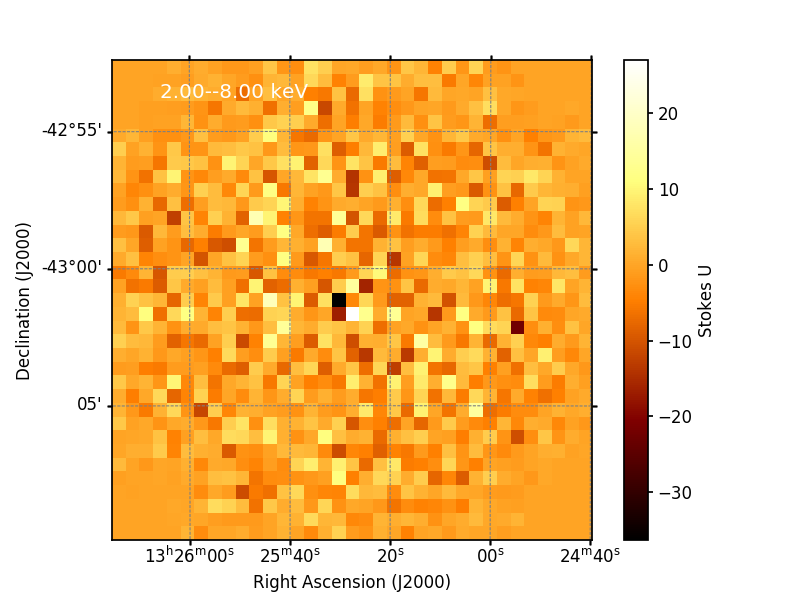}
\includegraphics[width=0.43\textwidth]{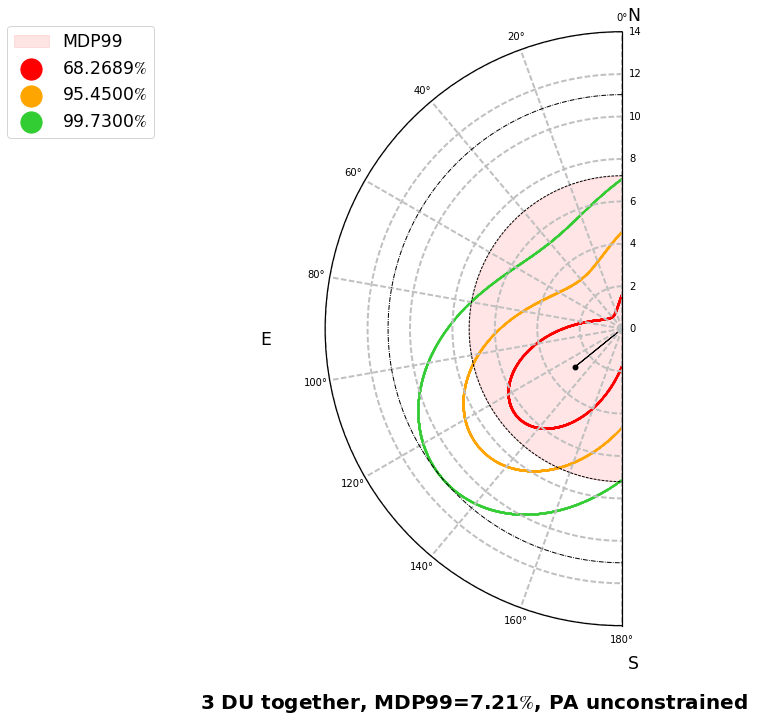}
\includegraphics[width=0.43\textwidth]{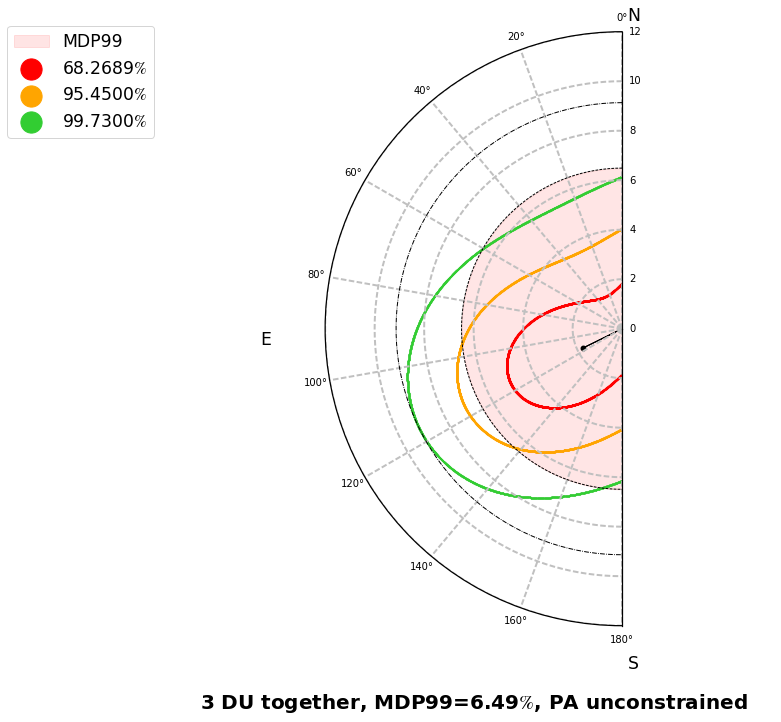}
\caption{Maps of the X-ray polarization in the core of Cen A. All of these sub-figures utilize the combined results of all three \textit{IXPE} detectors in the $2-8 \keV$ energy band. \textit{Top Left: } Normalized Stokes $Q$ within the core region. \textit{Top Right: } Normalized Stokes $U$ within the core region. \textit{Bottom Left: } Polar plot of the X-ray polarization in the core region, derived without event weighting. Red, orange, and green contours denote the $68.3\%$, \thinspace $95.5\%$, \thinspace and $99.7\%$ confidence intervals for the polarization degree and angle. The red shaded region is the MDP at $99\%$ confidence. \textit{Bottom Right: } Polar plot of the X-ray polarization derived using event weights. No statistically significant X-ray polarization is observed using either weighting method.}
\label{fig::PolMaps}
\end{figure}

Figure \ref{fig::PolMaps} presents the Stokes $Q$ and $U$ maps of the \textit{IXPE} field of view, along with polar plots showing the limits on the broad-band polarization of Cen A's core. In addition to the core region, two other sources are considered. Neither the X-ray jets and the ultraluminous X-ray source (ULX) near the core of Cen A have sufficient counts to measure statistically significant X-ray polarization. The broad-band polarization of the jet region has an MDP$_{99}$ of $40\%$, while MDP$_{99} = 80\%$ for the ULX region. Using the weighted polarization cubes gives MDP$_{99}$ values of $36\%$ and $73\%$  for the jet and ULX regions, respectively. These  values of MDP$_{99}$ all correspond to the optimistic assumption that all events within these regions originate from the sources and not the background. Given the low signal-to-noise (S/N) ratio of the imaging data, the polarization of these two sources is effectively unconstrained.
%LP changed
Using the likelihood base method that accounts for background (see \S~\ref{sec:ixpe}) from a region at comparable distance from the core but on the west side of it, the polarization of jet and ULX cannot be established even if they were fully polarized (MDPs are $100\%$).
%Using the likelihood method that accounts for background (see \S~\ref{sec:ixpe}) from a region at comparable distance from the %core but on the west side of it, the jet and ULX MDPs are $>100\%$.
 Fitting the broad band ($2-8 \keV$) polarization data of the Cen A core (without weighting), then combining the data from all three telescopes simultaneously, gives no statistically significant polarization signal above $\mathrm{MDP}_{99} = 7.2\%$. Incorporating the weights into the calculation provides a stricter limit of $\mathrm{MDP}_{99} = 6.5\%$. Dividing the \textit{IXPE} data into smaller energy bins also does not yield any statistically significant polarization in any of the energy bands investigated.

 %%%%%%%%%%%%%%%%%% FIG AA %%%%%%%%%%%%%%%%%%%%%%
 %%%%%%%%%%%%%%%%%%%%%%%%%%%%%%%%%%%%%%%%%%%%

\subsection{Spectral Fitting}

%From P.O. Petrucci concerning the X-ray emission

We have modelled the broad-band (0.5--400 keV) \textit{Swift + \textit{IXPE}\footnote{For this spectral fit, no polarization information from \textit{IXPE} was utilized, only the Stokes I parameter spectrum.} + NuSTAR + INTEGRAL}  X-ray spectrum  of the core of Cen A with several simple yet physically motivated models. The broad-band emission above 2 keV (the lower-energy bound of the \textit{IXPE} data) can be well described by a simple power-law continuum plus a Gaussian-shaped Fe K$\alpha$ emission line with moderate intrinsic absorption ($\sim 10^{23} \cm^{-2}$). There is no apparent evidence of a high-energy cutoff within the \textit{INTEGRAL} data out to 400~keV. The spectrum below 2~keV shows strong evidence for another spectral component beyond a single intrinsically absorbed power law. 
The presence of the Fe K$\alpha$ line and the absence of a statistically significant Compton hump suggest that the observed spectrum originates in reflection from an optically thin plasma. Unless otherwise noted, all uncertainties correspond to $90\%$ ($\Delta \chi^{2} = 2.706$) confidence intervals and all upper limits correspond to $99\%$ ($\Delta \chi^{2} = 6.635$) confidence.

\begin{figure}
\centering
\includegraphics[width=0.68\textwidth]{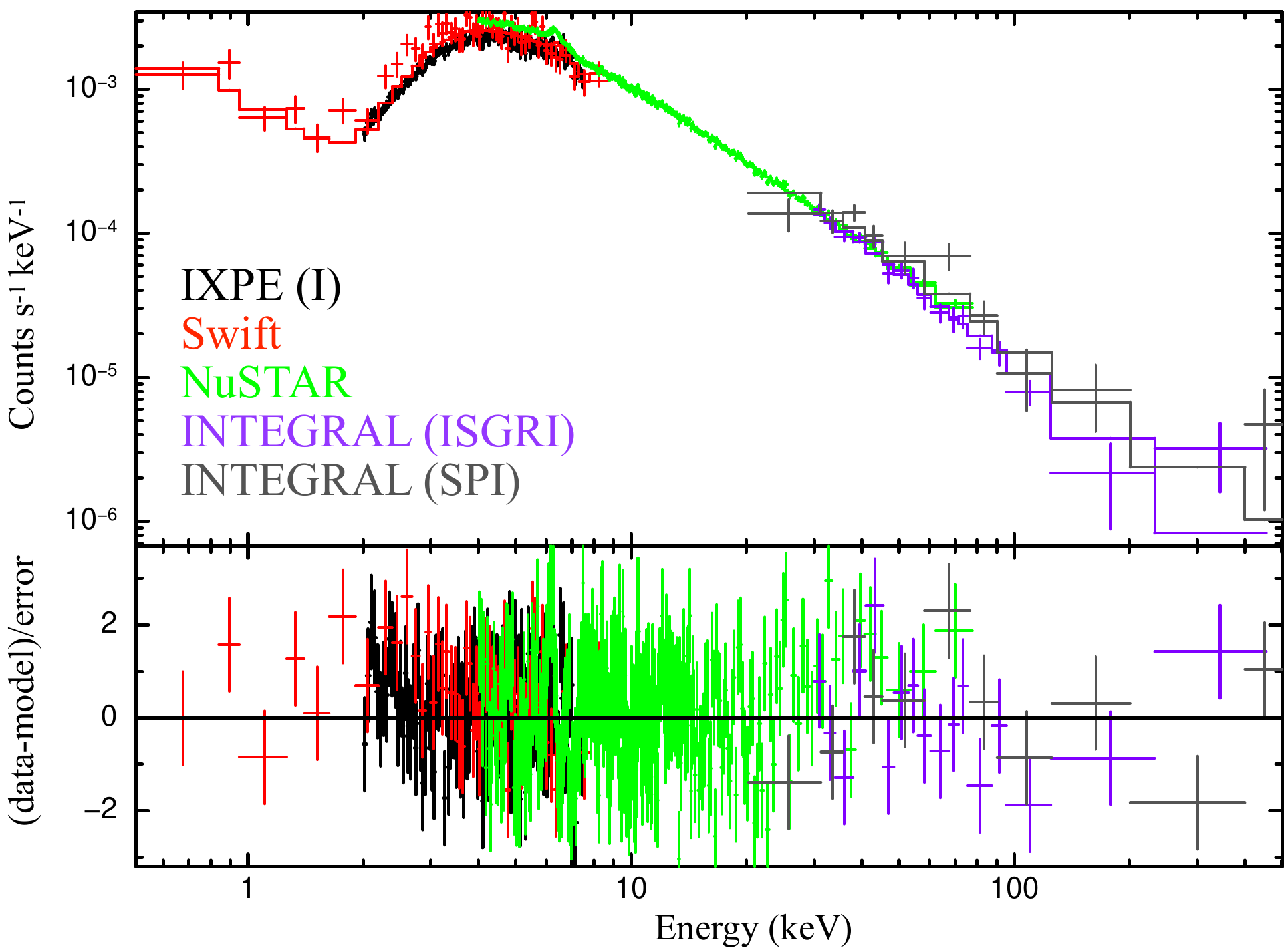}
\caption{Model fit of the joint \textit{Swift} (red),  \textit{IXPE} (black),  \textit{NuSTAR} (green), and  \textit{INTEGRAL} (purple and brown)  X-ray spectrum  of Cen A's core. This model corresponds to the simpler cutoff power-law model described in the text. No polarization information from \textit{IXPE} is used in this model fit.   }
\label{fig:CenASimpleSpectrum}
\end{figure}

Our primary model fit was performed using \texttt{XSPEC} version 12.12.1 \citep{Arnaud1996}, and assumes an intrinsically absorbed cutoff power law, a second power law with only Galactic absorption, and a Gaussian emission line (\texttt{const * tbabs * (ztbabs* (zcutoffpl + zgauss) + pow)} in \texttt{XSPEC}). We assume zero intrinsic width for the Fe K$\alpha$ line and a Galactic absorption column density of $N_{\rm H}= 2.36 \times 10^{20} \cm^{-2}$ \citep{HI4PI2016}. Our best-fit photon index for this model is $\Gamma = 1.88 \pm 0.03$. The best-fit intrinsic absorption column density is $1.48^{+0.05}_{-0.04} \times 10^{23} \cm^{-2}$, and the cutoff energy is $E_{\rm cut} > 500 \keV$ (the hard limit of this parameter in the \texttt{zcutoffpl} model in \texttt{XSPEC}). The cross-calibration constants for the various detectors differ from \textit{IXPE} Detector Unit 1 by $< 10\%$ for the other two \textit{IXPE} detectors, $\sim 30\%$ for the \textit{NuSTAR} FPMA/FPMB spectra, $\sim 10-15\%$ for the \textit{Swift} XRT spectra, and $\sim 40-60\%$ for the \textit{INTEGRAL} SPI/ISGRI spectra. The overall fit statistic of this model ($\chi^{2} = 1188$ for 1134 degrees of freedom) suggests that this model is a good fit to the data.  The resultant model fit and its residuals can be found in Figure \ref{fig:CenASimpleSpectrum}.

The line energy and flux of the Fe K$\alpha$ line are only slightly lower than the \textit{NuSTAR}-derived values quoted in \cite{furst16}. The line energy we measure is $E_{\rm Fe} = 6.33 \pm 0.04 \keV$ \citep[cf.\ 6.40 keV in][]{furst16} and the total flux of the line is $(2.27 \pm 0.23) \times 10^{-4} \thinspace \mathrm{ph} \cm^{-2} \s^{-1}$ \citep[c.f. $(2.88 \pm 0.22) \times 10^{-4} \thinspace \mathrm{ph} \cm^{-2} \s^{-1}$ in ][]{furst16}. We conclude that the flux of the Fe $K\alpha$ line has remained constant despite the factor of $\sim 3$ drop in the overall flux of the Cen A core. The measured Fe K$\alpha$ flux is also consistent with results from past observations using other X-ray telescopes \citep[e.g.][]{Grandi2003,Evans2004,markowitz07,Fukazawa2011}.

We added the second power-law component to fit the apparent residual counts observed at low energies ($\sim 0.3-2.0 \keV$) when only fitting the data with a single power-law component. The best-fit photon index for this second component is $\Gamma = 1.21^{+0.06}_{-0.18}$, and its normalization is approximately 2 orders of magnitude lower than the primary component. Compared to a simpler model with only the primary power-law component, including a second power law component reduces the total $\chi^{2}$ value of the fit by a highly significant amount ($\Delta \chi^{2} = -84$ with two fewer degrees of freedom). Converting this change in $\chi^{2}$ into a change in Bayesian Information Criterion ($\mathrm{BIC}$) for a model fit with 1150 data points and two additional parameters ($\Delta \mathrm{BIC} = 2 \ln{1150} + \Delta \chi^{2}$)  gives $\Delta \mathrm{BIC} = -69.9$, strongly favoring the presence of this second power law. The corresponding calculation for the Akaike Information Criterion (AIC) is $\Delta AIC = 4 + \Delta \chi^{2} = -80$, again validating the presence of this second power-law component.
%This second power-law 
This component appears to be physically separated from the primary power law, as the data strongly disfavor a model where this second power-law component is subject to same intrinsic absorption as the primary. We also attempted to fit these data to a similar model that replaces the second power law with a thermal \texttt{apec} model.  An \texttt{apec} component is well motivated by the observed thermal emission in the vicinity of Cen A that would reside within the \textit{Swift} point-spread function in projection observed by \textit{Chandra} \citep[e.g. ][]{Kraft2003}. The temperature and metallicity of this thermal component, however, cannot be well constrained by the spectral data. Fixing the temperature to $kT = 2.25 \keV$ and the metallicity to $0.4 Z_{\bigodot}$ with a free normalization parameter enables a fit with $\chi^{2}$ value statistically indistinguishable from our double power-law model.

As a further test of our spectral analysis, a {\sc Borus} component \citep[e.g.][]{Balokovic2018,Balokovic2019} was included to self-consistently model the neutral Compton reflection and Fe line. In {\sc Xspec} the model reads as: \texttt{constant} $\times$ \texttt{tbabs} $\times$ (\texttt{atable}(borus02\_v170323l.fits) + \texttt{atable}(borus02\_v170323k.fits) + \texttt{ztbabs} $\times$ \texttt{cutoffpl}) + \texttt{pow}). The reflector is assumed to have a toroidal geometry and the opening angle is fixed to $\Delta\Omega/(4\pi)=0.5$. %The best-fit model parameters and their corresponding $90\%$ confidence intervals can be found in Table \ref{tab::spectralfit}.  
The majority of the best-fit spectral parameters are broadly consistent with past measurements of Cen A obtained with \textit{NuSTAR} \citep[e.g.][]{furst16}. In particular, we find that the Compton reflection can be well explained in terms of scattering off Compton-thin material, with a column density log($N_{\rm H}/{\rm cm}^{-2})=23.08_{-0.15}^{+0.25}$. Similar to our simpler model, the BORUS modeling finds no evidence for a high-energy cutoff. At $90\%$ confidence the lower limit to the cutoff energy is $E_{\rm cut} > 590 \keV$, and $E_{\rm  cut} > 380 \keV$ at $99\%$ confidence. 

For both of the spectral model fits described above, the only parameter that strongly disagrees with the results of \cite{furst16} is the normalization of the primary continuum, which is consistently a factor of  $\sim 3-4$ times lower than that calculated in \cite{furst16}.
Such flux variations in this energy band are consistent with the factor of $\sim 3-4$ variation of the $20-100 \keV$ flux observed with the \textit{RXTE} over a 12.5 year period \citep{Rothschild2011}. As discussed in \cite{furst16}, the 2013 observation took place during a time where its flux was $\sim 40\%$ higher than the average flux of Cen A measured between 2003 and 2009 with \textit{INTEGRAL} \citep{beckmann11}.

%\begin{table}[htbp]
%\centering  
%%\begin{tabular}{cc}
%\hline \hline
%Parameter & Fit Value \\
%\hline 
%Cutoff power law \\ 
%$\Gamma$ & $1.85 \pm 0.01$ \\
%$E_{cut}$ ($\keV$) &  $> 590$ \\
%\hline 
%Intrinsic Absorption \\
%$\log_{10}\left({N_{H} /1 \cm^{-2}}\right)$ & $23.40 \pm 0.13$ \\ %
%\hline 
%Iron K Line \\
%$E_{Fe} (\keV)$ & $6.33^{+0.03}_{-0.04}$ \\
%Flux ($\mathrm{ph} \cm^{-2} \s^{-1}$) & $2.27 \times 10^{-4}$ %\\

%\hline \hline
%\end{tabular}
%\caption{\textbf{TODO: Update this table} A summary of the best-fit spectral parameters for Cen A from a broad-band fit of spectra extracted from \textit{Swift}, \textit{NuSTAR}, and \textit{INTEGRAL}. Describe goodness of fit and such other parameters.  }     
%\label{tab::spectralfit}   
%\end{table}

\subsection{Spectro-polarimetric Fitting}

 We utilized the combined data from \textit{IXPE}, \textit{NuSTAR}, and \textit{Swift} to perform a joint spectro-polarimetric fit to the Cen A core emission, using a simpler version of our cutoff power-law model. The only modification is that the redshifted cutoff power-law model component is replaced with a simple power law, as there is no ability to constrain the cutoff energy without the \textit{INTEGRAL} data. The polarization degree and angle are assumed to be constant with energy. In \texttt{XSPEC} this model is written as  \texttt{const $\times$ tbabs $\times$ (ztbabs $\times$ (polconst $\times$ pow + polconst $\times$ zgauss) +polconst $\times$ pow )}. As stated above, no weighting was applied to the events when constructing the \textit{IXPE} $I$, $Q$, and $U$ spectra. The polarization degrees of the Gaussian component and second power law were fixed to zero, meaning that our upper limits are strictly for the primary power-law component. 
 
 We present three versions of the spectral fit: one allowing the polarization angle $\psi$ to be fit as a free parameter, one where the polarization angle is fixed to $\psi = 55^{\circ}$ (parallel to the radio jets), and one where  $\psi = -35^{\circ}$ (perpendicular to the radio jets in the relevant \texttt{XSPEC} coordinate system). When the polarization angle is a free parameter, the $99\%$ confidence upper limit on the polarization degree is $\Pi < 8.0\%$. Fixing the polarization angles to $\psi = 55^{\circ}$ or $\psi = -35^{\circ} $ results in upper limits of $\Pi < 7.4 \%$ and $\Pi < 6.2 \%$, respectively. 
  All other fit parameters are statistically consistent with those from the simpler spectral modeling performed above, and the $\chi^{2}$ value of the fit (1259 for 1243 degrees of freedom) is in line with expectations for a ``good'' fit to the data.  
  
  \subsection{Summary of Results}
  We present the results of all six of the $99\%$ confidence upper limits on the X-ray polarization of Cen A in Table \ref{tab::MDPResults}. While all of these upper limits only differ from one another by a few percent, we choose to present the weighted value of MDP$_{99}$ as the ``primary'' upper limit for Cen A's polarization. This choice arises from two facts of the weighted analysis. The first reason is that the value of MDP$_{99}$ calculated with or without weights is independent of any assumptions or models for the SED. The second reason is that the weights have been designed and tested to increase the relative weight of events with reliable track reconstruction \citep{DiMarco2022}, therefore improving the sensitivity of the observation to polarization. This $\sim 10\%$ reduction in MDP$_{99}$ by using weights is in good agreement with expectations from pre-launch simulations \citep{Baldini2022}. 

\begin{figure}
\centering
\includegraphics[width=0.68\textwidth]{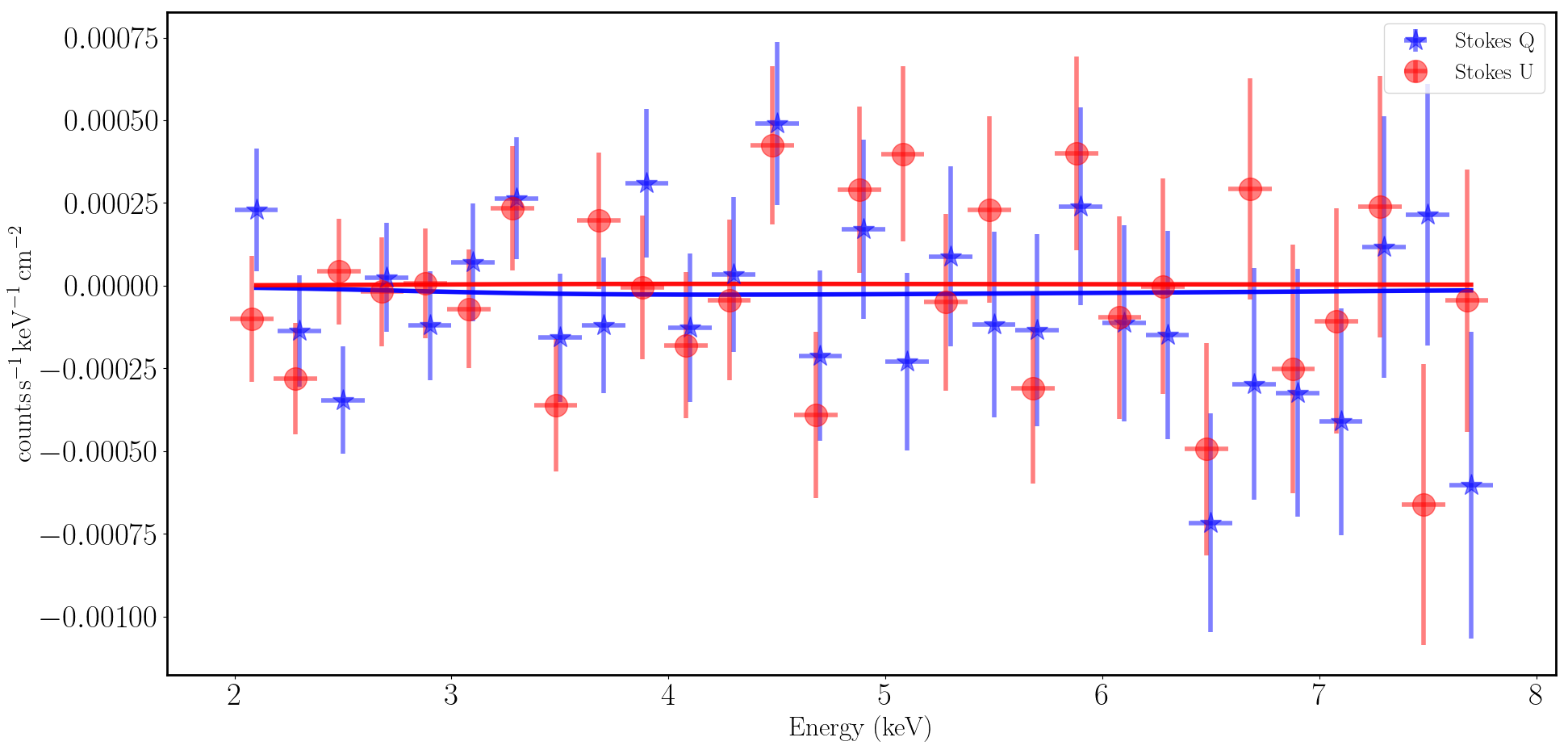}
\caption{Spectro-polarimetric model fit of the observed Stokes $Q$ (in blue) and $U$ (in red) spectra for Cen A. For presentation purposes only the $Q$ and $U$ spectra have been slightly offset in energies. This model assumes a constant polarization degree and angle for the primary power-law continuum and an unpolarized Fe K$\alpha$ line and secondary power law. This polarization model is determined simultaneously with the emission components (see main text for details).   No statistically significant polarization is measured with this fit.    }
\label{fig:CenAPolSpectrum}
\end{figure}

\begin{table}[]
    \centering
    \begin{tabular}{c c c c c}
    \hline
     Method & Background  & Weighting & $\psi$    & $99\%$ Limit \\
     \hline \hline
    PCUBE  & None & None & Free & $7.2 \%$ \\
    PCUBE  & None & N$_{\text{eff}}$ & Free & $6.5 \%$ \\
    Likelihood & In Calculation & None & Free & $7.1\%$ \\
    \hline
    Spectro-polarimetric & Subtracted & None & Free & $8.0\%$ \\
    Spectro-polarimetric & Subtracted & None & $55^{\circ}$ & $7.4\%$ \\
    Spectro-polarimetric & Subtracted & None & $-35^{\circ}$ & $6.2\%$ \\

    \hline

    \end{tabular}
    \caption{Summary of all calculations of the MDP$_{99}$ discussed in this paper. The PCUBE and Likelihood calculations are independent of any spectral model, while the spectro-polarimetric calculations are dependent on the best-fit Cen A spectral model. For the model-independent calculations, the $99\%$ limit is calculated as MDP$_{99}$, whereas for the model-dependent calculations we use the model fits to determine $99\%$ confidence upper-limits on the polarization degree parameter. }
    \label{tab::MDPResults}
\end{table}

\section{Discussion}

The apparent lack of a Compton hump suggests that the reflection component of the X-ray spectrum in Cen A arises primarily from reflection off an optically thin medium. On the other hand, the absence of any apparent cutoff in the power-law continuum agrees with past observations \citep[e.g.,][]{beckmann11,furst16}. 
This differs from the typical X-ray spectrum of Type 1 AGN \citep[e.g.][]{1997ESASP.382..373Z,2000ApJ...542..703Z,fab15,tor18,lan19} where a cut-off is generally interpreted as the signature of Compton scattering of seed photons in a thermal corona. 
Assuming that spectrum is produced by thermal Comptonization, we can estimate the lower limit to the electron temperature  from the lower limit on the cutoff energy as measured by \textit{INTEGRAL}, as $kT \approx E_c/2 > 200$~keV  \citep[e.g.][]{1995ApJ...449L..13S,pet01,mid19}. 
At such a high temperature, the Comptonization spectrum would deviate significantly from a single power law  \citep{coppi99}. 
This would suggest that thermal Comptonization is not the dominant process to produce the hard X-rays in Cen A and that non-thermal emission from the jet likely plays an important role.
The spectrum extending beyond 511 keV, the rest-mass energy of the electron, would be prone to photon-photon electron-positron pair-production which would affect the shape of the spectrum. 
Absence of any significant features the spectrum indicates that the source is not compact enough for this effect to play any role \citep{Svensson94}, adding further arguments in favor of the jet origin of the spectrum.

Variability of the different spectral components of Cen A between current and past observations \citep[in particular those of][]{furst16} can provide additional information about the structure of the X-ray emission and central black hole. The lack of variability in the Fe K$\alpha$ line \citep[see also e.g.][]{Rothschild2011}, despite the much higher variability in the continuum \citep[e.g.][]{Evans2004}, is indicative of a scenario where the reflecting medium is not only optically thin, but also quite distant from the region where the non-thermal X-rays are produced.

The upper limit to the degree of polarization measured by \textit{IXPE} provides important information about the particle population responsible for generating the X-rays. Given the low polarization degrees measured for Cen A in the radio and IR bands, however, it is not surprising that significant polarization was not measured in the \textit{IXPE} bandpass. Even using the optimistic assumption that the $11\%$ polarization degree at $2 \thinspace \mu m$ \citep{Capetti2000} originates entirely from synchrotron radiation (hence dismissing any polar scattering contribution), our measured limits for the X-ray polarization degree remain consistent with expectations from synchrotron self-Compton emission, where the polarization degree of the Compton scattered X-rays is predicted to be $\sim 2-5$ times smaller than that of the synchrotron radiation acting as seed photons for Compton scattering  \citep{1973A&A....23....9B,1994ApJS...92..607P,1994MNRAS.268..451C,Peirson2019}. We note that Compton scattering of unpolarized seed photons, such as the Cosmic Microwave Background or starlight from the host galaxy of Cen A \citep{Tanada2019} by isotropically distributed relativistic electrons in the jet would produce virtually unpolarized X-rays \citep{1970A&A.....7..292B,1993A&A...275..325N,1994ApJS...92..607P} because of relativistic aberration and resulting random rotation of polarization plane. 
Thus with a significantly longer exposure with \textit{IXPE}, where MDP$_{99}$ could reach levels of $\sim 1 \%$, the detection of polarization would be a strong argument in favor of synchrotron radiation as a source of seed photons for Compton scattering. The increase in exposure time would be significant, however, as MDP$_{99} \sim 4.29 /\sqrt{N}$ for the total number of observed source events $N$. Assuming the same count rate for the \textit{IXPE} observations presented here suggests a total of $\sim 5 \thinspace \mathrm{Ms}$ will be required to reach MDP$_{99} \sim 1 \%$. Even if future observations of Cen A were taken when the flux is a factor of $\sim 3$ higher, we would still need $\sim 1.6-2 \thinspace \mathrm{Ms}$ to reach such an MDP$_{99}$.
The observed limits on X-ray polarization are in tension with a physical scenario where the X-ray emission arises from hadronic jets. Processes involving hadrons in jets have been suggested as a possible source of the higher than expected emission of Cen A at TeV energies \citep{Abdo2010,Joshi2013}, and such models applied to blazars predict X-ray polarization degrees as high as $\sim 50-80\%$, much higher than our limits allow \citep{Zhang2013}.

All of the data presented in this work and in previous studies are consistent with a scenario where the majority of the X-rays observed in the core of Cen A originate from Compton scattering of lower-energy photons by non-thermal electrons, probably accelerated in regions within a few parsecs of the black hole. Since the polarization measurements for Cen A, spanning from radio to X-ray wavelengths, now sample both the synchrotron and Compton scattering components of the SED it remains plausible that the non-thermal electrons are accelerated in a region in the core with disordered magnetic fields.  The observed polarization of Cen A in the 2 -- 100 microns waveband \citep{Capetti2000,Lopez2021} likely arises from dichroic emission and absorption by aligned dust grains instead of direct emission from a non-thermal synchrotron component.  Future studies of other AGN with \textit{IXPE} will offer a clearer picture as to whether the low X-ray polarization degree observed in Cen A is typical of radio galaxies, or if it is an outlier among the radio galaxy population. 

\section*{Acknowledgments}
%Include acknowledgments of funding, any patents pending, where raw data for the paper are deposited, etc.

The Imaging X ray Polarimetry Explorer \textit{IXPE} is a joint US and Italian mission.  The US contribution is supported by the National Aeronautics and Space Administration (NASA) and led and managed by its Marshall Space Flight Center (MSFC), with industry partner Ball Aerospace (contract NNM15AA18C).  The Italian contribution is supported by the Italian Space Agency (Agenzia Spaziale Italiana, ASI) through contract ASI-OHBI-2017-12-I.0, agreements ASI-INAF-2017-12-H0 and ASI-INFN-2017.13-H0, and its Space Science Data Center (SSDC), and by the Istituto Nazionale di Astrofisica (INAF) and the Istituto Nazionale di Fisica Nucleare (INFN) in Italy.  This research used data products provided by the IXPE Team (MSFC, SSDC, INAF, and INFN) and distributed with additional software tools by the High-Energy Astrophysics Science Archive Research Center (HEASARC), at NASA Goddard Space Flight Center (GSFC).  We acknowledge Dawoon E. Kim for providing the polar-plot script. J.R. acknowledges financial support under the INTEGRAL ASI-INAF agreement 2019-35-HH.0 and ASI/INAF No. 2019-35.HH.0. The research leading to these results has received funding from the European Union's Horizon 2020 Programme under the AHEAD2020 project(grant agreement n. 871158).  The \textit{INTEGRAL} SPI project has been completed under the responsibility and leadership of CNES.  Part of the French contribution is supported by the Scientific Research National Center (CNRS) and the French spatial agency (CNES). We are grateful to ASI, CEA, CNES, DLR, ESA, INTA, NASA and OSTC for support. The research at Boston Unviversity was supported in part by National Science Foundation grant AST-2108622. The IAA-CSIC group acknowledges financial support from the Spanish "Ministerio de Ciencia e Innovaci\'on” (MCINN) through the “Center of Excellence Severo Ochoa” award for the Instituto de Astrof\'isica de Andaluc\'ia-CSIC (SEV-2017-0709) and through grants AYA2016-80889-P and PID2019-107847RB-C4
%Here you should list the contents of your Supplementary Materials -- below is an example. 
%You should include a list of Supplementary figures, Tables, and any references that appear only in the SM. 
%Note that the reference numbering continues from the main text to the SM.
% In the example below, Refs. 4-10 were cited only in the SM.  

%% For this sample we use BibTeX plus aasjournals.bst to generate the
%% the bibliography. The sample631.bib file was populated from ADS. To
%% get the citations to show in the compiled file do the following:
%%
%% pdflatex sample631.tex
%% bibtext sample631
%% pdflatex sample631.tex
%% pdflatex sample631.tex

\bibliography{CenARefs}
\bibliographystyle{aasjournal}

%% This command is needed to show the entire author+affiliation list when
%% the collaboration and author truncation commands are used.  It has to
%% go at the end of the manuscript.
\allauthors

%% Include this line if you are using the \added, \replaced, \deleted
%% commands to see a summary list of all changes at the end of the article.
%\listofchanges

\end{document}